\documentclass[a4paper,twoside]{article}

\usepackage{epsfig}
\usepackage{subfig}
\usepackage{calc}
\usepackage{amssymb}
\usepackage{amstext}
\usepackage{amsmath}
\usepackage{amsthm}
\usepackage{multicol}
\usepackage{pslatex}
\usepackage{apalike}
\usepackage{algorithm2e}
\usepackage[bottom]{footmisc}
\usepackage{SCITEPRESS}
\usepackage{booktabs}
\usepackage{float}
\usepackage{bm}
\usepackage[hidelinks]{hyperref}
\usepackage[capitalise]{cleveref}
\creflabelformat{equation}{#2#1#3}

\begin{document}

\title{Disentangling Quantum and Classical Contributions in Hybrid Quantum Machine Learning Architectures}


\author{\authorname{Michael Kölle\sup{1}, Jonas Maurer\sup{1}, Philipp Altmann\sup{1}, Leo Sünkel\sup{1}, Jonas Stein\sup{1} and Claudia Linnhoff-Popien\sup{1}}
\affiliation{\sup{1}Institute of Informatics, LMU Munich, Munich, Germany}
\email{michael.koelle@ifi.lmu.de}
}


\abstract{
    Quantum computing offers the potential for superior computational capabilities, particularly for data-intensive tasks. However, the current state of quantum hardware puts heavy restrictions on input size. To address this, hybrid transfer learning solutions have been developed, merging pre-trained classical models, capable of handling extensive inputs, with variational quantum circuits. Yet, it remains unclear how much each component -- classical and quantum -- contributes to the model's results. We propose a novel hybrid architecture: instead of utilizing a pre-trained network for compression, we employ an autoencoder to derive a compressed version of the input data. This compressed data is then channeled through the encoder part of the autoencoder to the quantum component. We assess our model's classification capabilities against two state-of-the-art hybrid transfer learning architectures, two purely classical architectures and one quantum architecture. Their accuracy is compared across four datasets: Banknote Authentication, Breast Cancer Wisconsin, MNIST digits, and AudioMNIST. Our research suggests that classical components significantly influence classification in hybrid transfer learning, a contribution often mistakenly ascribed to the quantum element. The performance of our model aligns with that of a variational quantum circuit using amplitude embedding, positioning it as a feasible alternative.
}

\onecolumn \maketitle \normalsize \setcounter{footnote}{0} \vfill

\section{INTRODUCTION} \label{sec:introduction}
In recent years, remarkable progress has been made in the field of machine learning, leading to breakthroughs in various areas such as image recognition \cite{dosovitskiy_image_2021} and speech recognition \cite{schneider_wav2vec_2019}. With the advancement of technology, the scale and complexity of data continue to increase, posing significant challenges for classical computational methods, including the curse of dimensionality \cite{bellman_dynamic_1957}: as the number of input features increases, the data required for accurate generalization grows exponentially, making it harder for classical algorithms to handle high-dimensional datasets effectively. In this context, the use of quantum computers promises performance advantages.

However, we are currently in the Noisy Intermediate-Scale Quantum (NISQ) era, characterized by not only a restricted number of qubits within the quantum circuit but also limitations on circuit depth and the quantity of operations conducted on the qubits \cite{preskill_quantum_2018}.

To overcome these limitations, approaches which combine classical neural networks (NN) with quantum circuits are subject to increased research. Mari et al. \cite{mari_transfer_2020} propose the Dressed Quantum Circuit (DQC), where a variational quantum circuit (VQC) is framed by a classical pre-processing NN and a classical post-processing NN. This approach is then combined with transfer learning. A major problem with a hybrid approach like this is the uncertainty of the actual contribution of the VQC to the classification performance and whether it provides any additional benefit over a purely classical NN as the VQC adds additional calculation time and complexity.

Another approach which builds on transfer learning is Sequential Quantum Enhanced Training (SEQUENT) \cite{altmann_sequent_2023}. Here, the post-processing layer is omitted and the training consists of two steps: classical and quantum. In the classical step, the model consists of a pre-processing layer and a surrogate classical classifier. This proxy is replaced by a VQC after pre-training and the corresponding quantum weights are optimized while the classical weights are frozen. Again, the influence of the respective classical and quantum parts of the model is ambiguous.

In this work, we propose an alternative approach to address the aforementioned challenges of the NISQ era. Instead of using transfer learning, the encoder part of an autoencoder (AE) is utilized to compress the input data into a lower-dimensional space. Subsequently, this reduced input is passed to a VQC, which then classifies the data. By using an AE for the compression, we aim to provide a more transparent understanding of the actual classification performance of the VQC, as the AE is solely trained on the reconstruction loss of the input data. The performance of our approach is then compared to a DQC, SEQUENT, a classical NN with the uncompressed and compressed input, and eventually with a pure VQC, which uses amplitude embedding. The models were trained and compared with each other on the datasets Banknote Authentication, Breast Cancer Wisconsin, MNIST, and AudioMNIST -- ranging from medical to image and audio data. Hence, our contributions can be summarized as follows:
\begin{itemize}
    \item We propose an alternative approach for handling high-dimensional input data in quantum machine learning (QML)
    \item We evaluate the individual performance of classical and quantum parts in hybrid architectures
\end{itemize}

\noindent All experiment data and code can be found here \footnote{https://github.com/javajonny/AE-and-VQC}. 

\section{VARIATIONAL QUANTUM ALGORITHMS} \label{sec:VQAandQML}

One of the most promising strategies for QML algorithms are variational quantum algorithms (VQA) \cite{cerezo_variational_2021}, which can be used e.g. as classifiers \cite{schuld_circuit-centric_2020,farhi_classification_2018}.
Generally, VQAs allow us to use quantum computing in the NISQ era by utilizing a hybrid approach of a quantum computer, but classical optimization strategies in an iterative quantum-classical feedback loop. A parametrized quantum circuit is used to change qubit states with different gate operations and a classical computer to optimize the parameters of the circuit. Hybrid in this context does not mean that a combined architecture with a classical and a quantum part is applied, but rather that solely a quantum circuit is used in combination with classical optimization strategies. The goal is to minimize a specified cost function $C$ in the training process by finding the optimal parameters $\theta^*$ for the quantum circuit \cite{abohashima_classification_2020,cerezo_variational_2021}:
$$\theta^* = \underset{\theta}{\operatorname{argmin}}\;C(\theta)$$
The first step in developing a VQA is defining a cost function $C$, parametrized by $\theta$:
$$C(\theta) = f(\{\rho_k\}, \{O_k\}, U(\theta))$$
with $f$ being a function determined by the concrete task, $U(\theta)$ being a parametrized unitary, $\theta$ being composed of discrete and continuous parameters, $\{\rho_k\}$ being input states from the training set and $\{O_k\}$ being a set of observables \cite{cerezo_variational_2021}.
The parameters and operations define the \textit{ansatz}, which is a parametrized quantum circuit -- a VQC. It can be formalized by the unitary $U(\theta)$, which is a product of $L$ unitaries sequentially applied \cite{cerezo_variational_2021}:
$$U(\theta) = U_L(\theta_L) \ldots U_2(\theta_2)U_1(\theta_1)$$
Each unitary of this product can further be decomposed into a sequence of unparametrized and parametrized gates:
$$U_l(\theta_l) = \prod_{m} e^{-i\theta_mH_m}W_m$$
$\theta_l$ is the $l$-th element in $\theta$, $e^{-i\theta_mH_m}$ is a parametrized gate with $H_m$ as a Hermitian operator, and $W_m$ represents an unparametrized unitary \cite{cerezo_variational_2021}. 
We will use a layered gate ansatz \cite{schuld_circuit-centric_2020}, where a sequence of gates is repeated and is therefore a hyperparameter. Mari et al. \cite{mari_transfer_2020} and Altmann et al. \cite{altmann_sequent_2023} use angle embedding for mapping the classical data vector $x = (x_1, \ldots, x_\eta)$ to a state $|x\rangle$ with $\eta$ qubits in a quantum Hilbert space. The embedding of the classical vector can be formalized in the following way \cite{mari_transfer_2020}: 
 $$\mathcal{E} : x \rightarrow |x\rangle$$
After that, one or several entangling layers are applied, which consist of controlled-NOT gates (CNOT) and single-qubit rotations \cite{altmann_sequent_2023}. In order to obtain the classical output vector $y$ from the quantum circuit, the expectations values of $n_q$ observables $\hat{y} = (\hat{y}_1, \hat{y}_2, \dots, \hat{y}_{n_q})$ must be calculated. In summary, the measurement layer maps from the quantum state back to a classical output vector $y$ \cite{mari_transfer_2020}:
$$\mathcal{M}: |x\rangle \rightarrow y = \langle x|\hat{y}|x\rangle$$
The whole VQC is therefore defined by (c.f. \cite{mari_transfer_2020}):
$$Q = \mathcal{M} \circ U(\theta) \circ \mathcal{E}$$
where $Q$ stands for the complete circuit. A classification task can be successfully solved with the proposed architecture \cite{mari_transfer_2020}.
Once the cost function and the VQC are defined, the next step is the training process, where parameters of the ansatz are optimized according to an objective function. For this purpose, one can use common classical optimization algorithms like Stochastic Gradient Descent, which are executed on classical computers. To compute the necessary gradients, the \textit{parameter shift rule} can be used to calculate the partial derivatives \cite{mitarai_quantum_2018,schuld_evaluating_2019}. 
\section{RELATED WORK} \label{sec:related-work}

VQCs have proven their potential at solving classification tasks in the context of supervised learning \cite{mitarai_quantum_2018,schuld_circuit-centric_2020,mari_transfer_2020,altmann_sequent_2023} and are therefore used in our approach. In order to use a VQC with higher-dimensional input in the NISQ era, transfer learning in hybrid classical-quantum networks is often applied. Mari et al. \cite{mari_transfer_2020} propose the DQC, which consists of a classical pre-processing and a post-processing part, framing a VQC. Additionally, they apply different transfer learning architectures. According to the authors, the most appealing one is classical to quantum transfer learning where a classical pre-trained NN extracts features, which are subsequently passed to a variational quantum classifier.

Altmann et al. \cite{altmann_sequent_2023} also use transfer learning in their approach SEQUENT, as the training takes place in a two-step process. The architecture consists of a classical compression layer and a classification component, whereas the latter is embodies by a classical NN in the first training round and gets replaced by a VQC in the next training step. Therefore, SEQUENT implements sequential training through the two-step process -- unlike the DQC, where all components are trained concurrently \cite{altmann_sequent_2023}. The second difference compared to a DQC is that SEQUENT omits the post-processing part.

Both models have in common that the influence of the VQC in the classification task cannot be exactly quantified and assessed. It may be, that the classical NN already does all or most of the classification in the pre-processing. Our model, which uses an AE, addresses precisely this problem because the input is only compressed to a lower dimension, not already classified, as our results also show.
\section{OUR APPROACH} \label{sec:approach}

In this chapter, we introduce our approach. First, we will describe the architecture of the AE and the VQC individually. Subsequently, we will illustrate how these two components are combined to achieve the desired reduction in dimensionality.

\subsection{Autoencoder for Dimensionality Reduction}\label{subsec:AutoencoderDimensionalityReduction}
An AE is a special kind of NN that can be used to encode the input into a reduced, but still meaningful representation, and therefore acts as a feature extractor \cite{goodfellow_deep_2016}. The idea is to overcome the limitations of VQCs in the current NISQ era by reducing the input dimension to the smallest meaningful dimension -- the number of output classes. Therefore, the number of qubits in the VQC only depends on the number of output classes, which typically is much lower than the number of input features. The reduced input can then be fed into the VQC. The proposed AE has the following architecture. 

\subsubsection{Encoder}\label{subsubsec: EncoderAE}
We define an adaptable approach that works for diverse datasets with different characteristics. Each layer halves the dimension of the preceding layer by reducing the number of neurons with floor division. This process is repeated until a layer has the same number of neurons as output classes. If there are less neurons than output classes, the number of neurons in the last layer is set directly to the output classes. To enable non-linear transformations, the Rectified Linear Unit (ReLU) activation function has proven its effectiveness \cite{nair_rectified_2010,krizhevsky_imagenet_2017}:
$$ReLU(x) = max(0,x)$$
Due to its simplicity and reliably good performance in NNs \cite{ramachandran_searching_2017}, we use ReLU in this AE architecture after the input layer and between the hidden layers. 
After the last layer, on the other hand, a Sigmoid function is used as an activation function, which converts any input to the range $[0,1]$ and therefore aligns with our pre-processed input data:
$$S(x) = \frac{1}{1+e^{-x}}$$ 

\subsubsection{Decoder}\label{subsubsec: DecoderAE}
The aim of the decoder is the reconstruction of the input from the compressed representation with minimal reconstruction loss. It corresponds to the mirrored architecture of the encoder and therefore has the same number of layers. Again, after each layer the ReLU activation function is applied -- except for the last layer where Sigmoid is used to ensure that the reconstructed input is in the same range as the original input after the pre-processing.

\subsection{Variational Quantum Circuit}\label{subsec:VariationalQuantumCircuit}
In the following, we will explain the architecture of the proposed VQC, which acts as a classifier. It consists of the three components: state preparation, entangling layers, and the measurement layer.

\subsubsection{State Preparation}\label{subsubsec: StatePreparation}
At first, the provided input in the range $[0,1]$ is converted to the range $[-\frac{\pi}{2}, \frac{\pi}{2}]$ to be within the range of typical angles used in quantum gates. Our presented VQC then uses angle embedding to encode the feature vector of dimension $N$ into the rotation angles of $n$ qubits, with $N \leq n$, in the quantum Hilbert space.

Initially, qubits are set to the computational basis state $|0\rangle$. In the first step, a layer of single-qubit Hadamard gates is applied to each qubit in the circuit and hereby transforms the basis state $|0\rangle$ to an equal superposition $|+\rangle$, making the quantum state unbiased with respect to $|0\rangle$ and $|1\rangle$: 
$$|+\rangle = \frac{1}{\sqrt{2}} |0\rangle + \frac{1}{\sqrt{2}} |1\rangle$$
In the second step, a Ry-gate is applied to each qubit, which performs the actual angle embedding.

\subsubsection{Entangling Layers}\label{subsubsec: VariationalLayers}
To save time in the optimization and training process by reaching the convergence point of the accuracy faster, weight remapping is applied \cite{kolle_improving_2023,kolle_weight_2023}. For angle embedding, the following function constitutes the largest improvement:
$$\varphi(\theta) = 2 \cdot arctan(2 \cdot \theta)$$
The entangling layers apply a sequence of trainable operations to the prepared states, whereby the general architecture is inspired by the \textit{model circuit} \cite{schuld_circuit-centric_2020} and the architecture of SEQUENT \cite{altmann_sequent_2023}. Each of the entangling layers consists of a CNOT ladder resulting in entanglement between qubits, followed by Ry-gates that apply parametrized rotations around the y-axis.

\subsubsection{Measurement Layer}\label{subsubsec: Measurement}
The last component performs the measurements of each wire in the computational basis. More specifically, the expectation value of the Pauli-Z operator is calculated and returned for each wire.

\subsection{Integration of the Autoencoder and the Variational Quantum Circuit}\label{subsec:CombinedModel}

Here we describe how the AE and VQC are connected.

The first step consists of data pre-processing, which includes one-hot encoding of the labels and the normalization of the input features to the range $[0, 1]$. Following this, an AE with the just presented architecture has to be initialized and trained in order to reduce the reconstruction loss. We selected the mean square error as the loss function and the Adam optimizer algorithm \cite{kingma_adam_2015} as the optimizer. After training the AE, only the optimized encoder is used by passing the input data to it and compressing the input dimension to the number of labels.

\begin{figure*}[hpbt]
  \centering
  \includegraphics[width=0.85\textwidth]{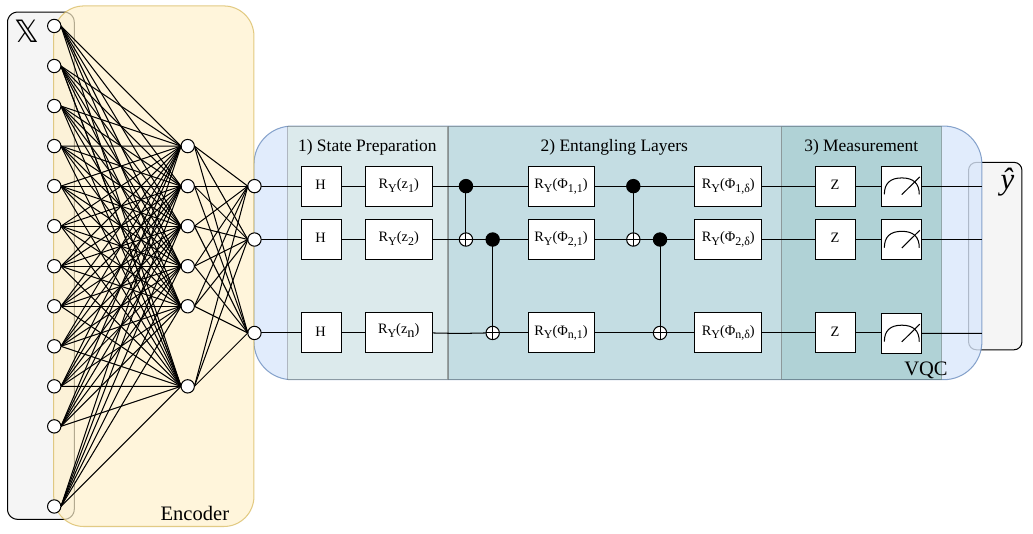}\\
  \caption[Architecture of our approach.]{Architecture of our approach, which consists of an encoder and a VQC parametrized with $\phi$. The input data is given by $\mathbb{X}$ and the prediction targets by $\hat{y}$. The VQC has $n$ qubits and consists of $\delta$ entangling layers.}
  \label{fig:Picture_Model_Architecture}
\end{figure*}

In order to classify the reduced features, the VQC must then be initialized. Thereby, for the number of wires in the circuit, the dimension of the compressed input is selected. After the initialization of the VQC follows the training of the same, for which classical optimization techniques can be used in an iterative process with a quantum-classical feedback loop. The Cross Entropy Loss is taken as the criterion and the Stochastic Gradient Descent for the optimization. Subsequently, the trained VQC is used for classification tasks. An illustration of our proposed architecture can be found in \cref{fig:Picture_Model_Architecture}.

\section{EXPERIMENTAL SETUP} \label{subsec:experimental-setup}

In this chapter, the datasets, baselines, and the hyperparameter optimization are presented. For the AE, Mean Square Error (MSE) as the reconstruction loss is the main evaluation metric. For our model and the baselines, the Cross Entropy Loss and especially the accuracy are relevant.

\subsection{Datasets and Pre-Processing}\label{subsec:Datasets}
The four different datasets Banknote Authentication, Breast Cancer Wisconsin, MNIST and AudioMNIST were chosen to demonstrate the versatility of the suggested approach. Each dataset is split into a training set, a validation set and a test set. Because the used datasets take a variety of different forms, we created custom datasets for each of them to take into account the specific data preparation. In this chapter, we introduce these datasets (summarized in \cref{tab:DatasetsOverview}).

\subsubsection{Banknote Authentication}\label{subsubsec:DatasetBanknoteAuthentication}
The first presented dataset is the \textit{Banknote Authentication} dataset for binary classification of genuine and forged banknotes \cite{lohweg_banknote_2013}. In total, it contains 1372 instances, 762 for the genuine class and 610 for the forged class. Each instance has 4 features.

In order to use the dataset, the data was split into two parts: the feature vectors and the labels. The latter was converted to integer labels and subsequently one-hot encoding was applied. Also the data was modified and scaled to the range $[0,1]$. After that, the custom dataset was split into a training, validation and testing set with a split of $8 : 1 : 1$.

\subsubsection{Breast Cancer Wisconsin}\label{subsubsec:DatasetBreastCancerWisconsin}
Another dataset used for binary classification is the \textit{Breast Cancer Wisconsin} dataset \cite{wolberg_breast_1995}. It contains 569 instances, 357 instances for the benign class and 212 for the malignant class. Each instance consists of 30 features, which describe the properties of the cell nuclei that are present in an image.

For the Breast Cancer Wisconsin dataset, the data pre-processing steps are similar to Banknote Authentication. The dataset can be split into data and labels. Since the labels were already available as integers, they can directly be one-hot encoded. As with Banknote Authentication, the data was scaled to a range of $[0,1]$ and then the dataset was split into a training, validation and testing set with a split of $8 : 1 : 1$.

\subsubsection{MNIST}\label{subsubsec:DatasetMNIST}
In order to demonstrate the versatility of the suggested approach we use the image dataset Modified National Institute of Standards and Technology (\textit{MNIST}) \cite{lecun_gradient-based_1998}. It contains 60,000 instances for training and 10,000 for testing. Each instance represents a grayscale image of a handwritten digit and is uniformly sized to $28\times28$ pixels -- i.e. 784 pixels per image\footnote{Flattening the input leads to a dimension of 784 for each vector.}. Each pixel takes a value from zero (black) to 255 (white), representing the intensity, i.e. the brightness or darkness. In addition, each image has a corresponding label indicating the true value from zero to nine.

The MNIST dataset can be imported as a separate train and a test dataset. Subsequently, the former was further split into a training and a validation set, whereby the validation set should have the same size as the test set. Therefore, the split was $\frac{5}{7} : \frac{1}{7} : \frac{1}{7}$.
Since every instance was sized uniformly to $28\times28$ pixels, the next step included flattening the samples into a one-dimensional tensor of size $784$. As before, one-hot encoding was applied to the labels. The last step included normalizing the data instances from the range $[0, 255]$ to $[0, 1]$ and converting it to the same data type as the labels.  

\subsubsection{AudioMNIST}\label{subsubsec:subDatasetAudioMNIST}
Becker et al. \cite{becker_interpreting_2019} present an audio dataset, which is accessible via the following link\footnote{https://github.com/soerenab/AudioMNIST}. It consists of 30,000 audio recordings in total and is called \textit{AudioMNIST}. The digits are spoken in English, containing 50 repetitions per digit for the 60 different speakers. The recordings are stored in the Waveform Audio File (WAV) format and the sampling rate is 48kHz. 

First, the data was split into training, validation, and testing lists with a split of $8 : 1 : 1$. We converted the audio files to Mel spectrograms and then to images. To enable comparability with the MNIST dataset, we reshaped the images to a dimension of $28\times28$. In the following step the tensors were flattened into a one-dimensional tensor of size $784$. The labels were one-hot encoded and the tensor with the data instances was normalized to the range $[0, 1]$.

\begin{table}[htbp]
\centering
\resizebox{\columnwidth}{!}{%
\begin{tabular}{l c c c}
\toprule
Dataset & \#Features & \#Classes & \#Samples\\
\midrule
Banknote Authentication & 4 & 2 & 1372 \\
Breast Cancer Wisconsin & 30 & 2 & 569 \\
MNIST & 784 & 10 & 70000 \\
AudioMNIST & 784 & 10 & 30000 \\
\bottomrule
\end{tabular}
} 
\caption[Dataset characteristics overview.]{Characteristics of the datasets Banknote Authentication, Breast Cancer Wisconsin, MNIST, and AudioMNIST.\label{tab:DatasetsOverview}}
\end{table}

\subsection{Baselines}\label{subsec:Baselines}
Our presented model will be compared against the following baselines: a VQC with amplitude embedding, a DQC, SEQUENT, a classical NN, and an AE in combination with a classical NN.

\subsubsection{Variational Quantum Classifier (Amplitude Embedding)}\label{subsubsec:BaselinesVQC}
An alternative approach for the dimensionality reduction is using a VQC with amplitude embedding. The architecture of the VQC just differs in the state preparation - we use the same architecture and number of entangling layers as in our model.
Amplitude embedding encodes the features of the input vector into the amplitudes of the qubits and makes superposition its advantage: an exponential number of features, $2^n$, can be mapped into the amplitude vector of $n$ qubits \cite{schuld_supervised_2018}. It's important to note that, for cases where $n$ is less than $\log_2N$, where $N$ represents the number of classical features, padding is applied to the original input, i.e. fill the vector with zeros. An additional requirement is the normalization of the padded input to unit length: $x_{padded}^T \cdot x_{padded} = 1$.

\subsubsection{Dressed Quantum Circuit}\label{subsubsec:BaselinesDressedQuantumCircuit}
This baseline is the DQC as suggested by Mari et al. \cite{mari_transfer_2020}. It consists of three parts. A classical pre-processing part, a VQC, and a post-processing part. The pre-processing part is a NN, consisting of one layer, which in our case reduces the input vector directly to the number of output classes, followed by applying a Sigmoid activation function. This reduced input is then passed to a VQC which has the same architecture and number of layers as our reference model. The post-processing part also consists of one layer and maps the data from the dimension of the VQC width to the number of output classes. In our case, both dimensions are the same.

The training consists of two stages. In the classical stage, just the weights of the pre- and post-processing are optimized. In the second stage -- the quantum stage -- these classical weights are frozen and just the parameters in the VQC are optimized. In the original paper, Mari et al. \cite{mari_transfer_2020} propose a number of schemes for transfer learning, with classical to quantum being the most appealing because one can use pre-trained NNs for image classification. 

We use this model as a baseline since the pre-processing layer also reduces the input. Here the classical pre-processing layer already classifies. The disadvantage of the DQC is that the effect of the classical and quantum parts are not separately assessable.

\subsubsection{SEQUENT}\label{subsubsec:BaselinesSEQUENT}
The third baseline SEQUENT was proposed by Altmann et al. \cite{altmann_sequent_2023}.
It consists of a classical compression layer and a classification part. The training takes place in two stages. In the classical stage, the classification part is a classical surrogate feed forward NN. We reduce the input directly to the number of output classes in the classical compression layer and then apply a Sigmoid activation function. This reduced input is then passed to the second NN for classification. The two parts are both trained and the weights are optimized.

In the quantum training step, the classical weights are frozen and the classical surrogate classification network is replaced by a VQC. For the training, just the quantum parameters are optimized. The VQC is defined as in the DQC. Altmann et al. \cite{altmann_sequent_2023} argue that this two-step procedure in SEQUENT and DQC can be seen as transfer learning because it is transferred from classical to quantum weights. It should be noted, however, that this does not correspond to the common definition of transfer learning, which refers to transferring knowledge between two different domains and tasks \cite{tan_survey_2018}. This model also has the same disadvantage as the DQC in that the effect of the classical and quantum components are not separately assessable.

\subsubsection{Classical Feed Forward Neural Network on Uncompressed Input}\label{subsubsec:BaselinesClassicalFNNonOriginalInput}
Another baseline is a classical feed forward NN which is introduced to verify if our model achieves a quantum speedup. The NN for this paper consists of one input layer with a variable number of nodes -- depending on the input dimension of the data. It is followed by a ReLU activation function and a hidden layer. This is followed by an output layer, which has the same number of nodes as output classes. The choice for the number of neurons in the hidden layer is determined by the total number of trainable parameters in our proposed approach with the AE and VQC, minus the neurons in the input layer $N_{input}$ and output layer $N_{output}$ of the NN. The formula for the total number $N_{total}$ of parameters in a network with one hidden layer is:

\begin{equation}
    \begin{split}
    N_{total} =& (N_{input} \cdot N_{hidden}) + N_{hidden} \\ 
               & + (N_{hidden} \cdot N_{output}) + N_{output}
    \end{split}
\end{equation}
The first term corresponds to the number of weights connecting the input and hidden layer. The second term is the number of bias parameters for the hidden layer. The third term is again the number of weights between the hidden and output layer. The last term embodies that every neuron in the output layer has its own bias.

By transformations we obtain the required number of neurons in the hidden layer $N_{hidden}$. For fairness reasons, this will be rounded up and at least one neuron is required:
$$N_{hidden} = \max\left(\left\lceil \frac{N_{total} - N_{output}}{N_{input} + N_{output} + 1} \right\rceil, 1\right)$$
$N_{total}$ is determined by our approach and corresponds to the total number of trainable parameters in the AE and the VQC.

\subsubsection{Classical Feed Forward Neural Network on Compressed Input}\label{subsec:BaselinesClassicalFNNonCompressedInput}
The last baseline is an AE which compressed the input data, in combination with a NN. The AE is the same as for our introduced model. The classical NN has the same architecture as the one just presented in \cref{subsubsec:BaselinesClassicalFNNonOriginalInput}. The only difference is the number of neurons in the hidden layer, especially how $N_{total}$ is obtained. Since the same AE is used as in the reference model, only the number of trainable parameters in the VQC is used for $N_{total}$ to obtain approximately the same number of parameters.

\subsection{Optimization, Training and Hyperparameters}\label{subsec:TrainingAndHyperparameters}
In this chapter we discuss the optimization, training process, and the obtained hyperparameters for each model. The applied optimization technique was grid search and all experiments were performed several times for multiple seeds to obtain a more reliable and robust result. Python (version 3.8.10) with the frameworks PyTorch (version 1.9.0+cpu) and PennyLane (version 0.27.0) were used for all of our experiments. For all plots in this paper, the exponential moving average with a smoothing factor $\alpha = 0.6$ was used to display the curves.

\subsubsection{Optimization of the Autoencoder}\label{subsubsec:OptimizationAutoencoder}
For the grid search of the AE, batch sizes of 32, 64, 128, 256 and for the learning rate the values 0.1, 0.01, 0.001, 0.0005 or 0.0001 were considered as possible hyperparameters. Each experiment was repeated five times per dataset, where the seeds took values from zero to four.

The MSE acted as the loss function and Adam as the optimizer. In order to allow all combinations to achieve convergence in validation loss, the number of epochs was set to 500 for all data sets. The determining metric for selecting the best hyperparameters per dataset was the mean test reconstruction loss. A summary of the best hyperparameter combinations can be found in \cref{tab:AEOptimization}.

\begin{table}[htbp]
\centering
\resizebox{\columnwidth}{!}{%
\begin{tabular}{l c c c c c c c}
\toprule
Dataset & Epochs & Learning Rate & Batch Size & Test Loss\\
\midrule
Banknote Authentication & 500 & 0.1 & 128 & $0.0046\pm0.0002$ \\
Breast Cancer Wisconsin & 500 & 0.01 & 32 & $0.0077\pm0.0034$ \\
MNIST & 500 & 0.001 & 64 & $0.0198\pm0.0012$ \\
AudioMNIST & 500 & 0.001 & 128 & $0.0006\pm0.0001$ \\
\bottomrule
\end{tabular}
} 
\caption[AE optimization with best hyperparameter combination for all datasets.]{AE optimization with the test reconstruction loss and $95\%$ confidence interval for the best hyperparameter combination per dataset.\label{tab:AEOptimization}}
\end{table}

\begin{table*}[tbp]
\centering
\resizebox{\textwidth}{!}{%
\begin{tabular}{l c c c c c c c}
\toprule
Model & AE+VQC (angle) & VQC (amplitude) & DQC & SEQUENT & AE+NN & NN \\ Dataset\\
\midrule
Banknote Authentication & $0.7841\pm0.0515$ & $0.8319\pm0.0321$ & $0.9928\pm0.0064$ & $0.9812\pm0.0197$ & $0.6884\pm0.1504$ & $0.9928\pm0.0090$  \\
Breast Cancer Wisconsin & $0.9018\pm0.0699$ & $0.8456\pm0.0520$ & $0.9579\pm0.0365$ & $0.9614\pm0.0358$ & $0.9298\pm0.0576$ & $0.9684\pm0.0542$ \\
MNIST & $0.5302\pm0.0704$ & $0.4472\pm0.0369$ & $0.8945\pm0.0107$ & $0.4992\pm0.1040$ & $0.8081\pm0.0771$ & $0.9853\pm0.0012$\\
AudioMNIST & $0.2239\pm0.0376$ &  $0.2651\pm0.0246$ & $0.4292\pm0.2978$ & $0.3999\pm0.0796$ & $0.2956\pm0.0498$ & $0.8759\pm0.0922$\\
\bottomrule
\end{tabular}
} 
\caption[Test accuracies for best hyperparameter combination.]{Test accuracies for the best combination of hyperparameters for each model and dataset.\label{tab:HyperparamOptimizationTable}}
\end{table*}

\subsubsection{Training Procedure}\label{subsubsec:OptimizationModels}
With the results for the hyperparameter optimization of the AE we can continue with the optimization of the models. Grid search was applied here as well, but just the learning rate was optimized. It could take the values 0.1, 0.01, and 0.001. In order to allow for a more direct comparison of the models, the batch size was set to five for all. The number of layers of the VQC was set to six. For the classical feed forward NN on the (un)compressed input, one hidden layer is considered. Each experiment was repeated for five seeds, ranging from zero to four.

For all models and experiments the Cross Entropy Loss was chosen as loss function and the Stochastic Gradient Descent as optimizer. The number of epochs was set to 100 for Banknote Authentication and Breast Cancer Wisconsin. Because of the longer computing time for MNIST and AudioMNIST, the number of epochs was set to 50 for the model training on these datasets -- this was sufficient to achieve convergence (c.f. \cref{fig:multipic}). It should be noted that the hybrid transfer learning models DQC and SEQUENT have been trained for the double amount of epochs because of the two training stages. The number of epochs for each stage was not halved because also the AE has been trained (for 500 epochs) until convergence. The crucial metric for selecting the best learning rate was the mean test accuracy over the five seeds. The best learning rate for each model and dataset can be seen in \cref{tab:BestLearningRateTable} and the achieved accuracies can be found in \cref{tab:HyperparamOptimizationTable}.

\begin{table}[htbp]
\centering
\resizebox{\columnwidth}{!}{%
\begin{tabular}{l c c c c c c c}
\toprule
Model & AE+VQC (angle) & VQC (amplitude) & DQC & SEQUENT & AE+NN & NN \\ Dataset\\
\midrule
Banknote Authentication & $0.01$ & $0.01$ & $0.1$ & $0.1$ & $0.1$ & $0.1$ \\
Breast Cancer Wisconsin & $0.1$ & $0.01$ & $0.1$ & $0.1$ & $0.1$ & $0.1$ \\
MNIST & $0.01$ & $0.01$ & $0.01$ & $0.001$ & $0.01$ & $0.1$ \\
AudioMNIST & $0.001$ & $0.1$ & $0.1$ & $0.1$ & $0.01$ & $0.1$ \\
\bottomrule
\end{tabular}
} 
\caption[Optimal learning rates for models.]{Optimal learning rate for each model and dataset obtained from the grid search.\label{tab:BestLearningRateTable}}
\end{table}

\section{RESULTS} \label{sec:results}

We selected the obtained hyperparameters from the optimization process (see \cref{tab:BestLearningRateTable}) based on the best performance for each model per dataset and ran each experiment ten times with different seeds (ranging from zero to nine). The number of layers for the VQC stayed at six, and the batch size for training was kept at five. The models were trained for the same number of epochs as in the optimization.

\begin{figure*}[htb]
 \centering
  \subfloat[Banknote Authentication]{
   \label{fig:multipic:Banknote} 
   \includegraphics[width=0.48\linewidth]{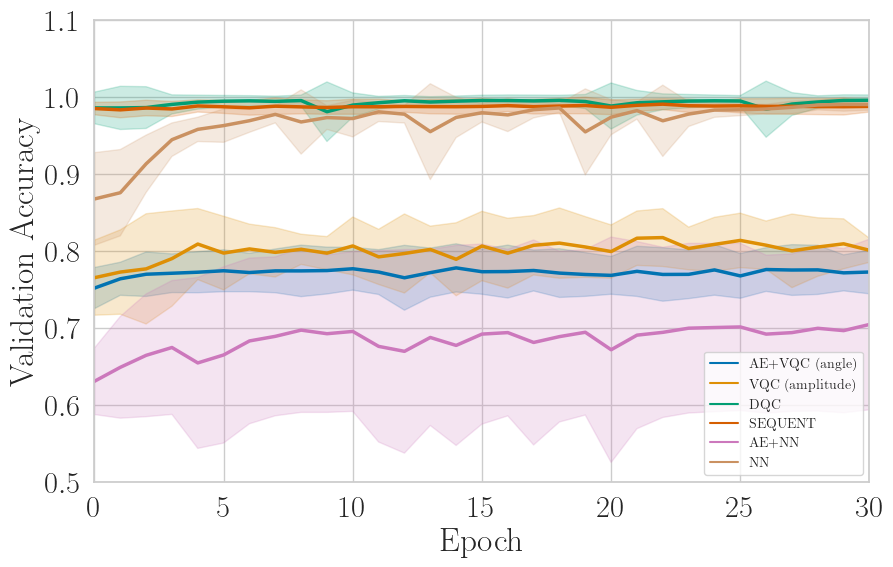}}
  \hfill
  \subfloat[Breast Cancer Wisconsin]{
   \label{fig:multipic:BreastCancerDetection} 
   \includegraphics[width=0.48\linewidth]{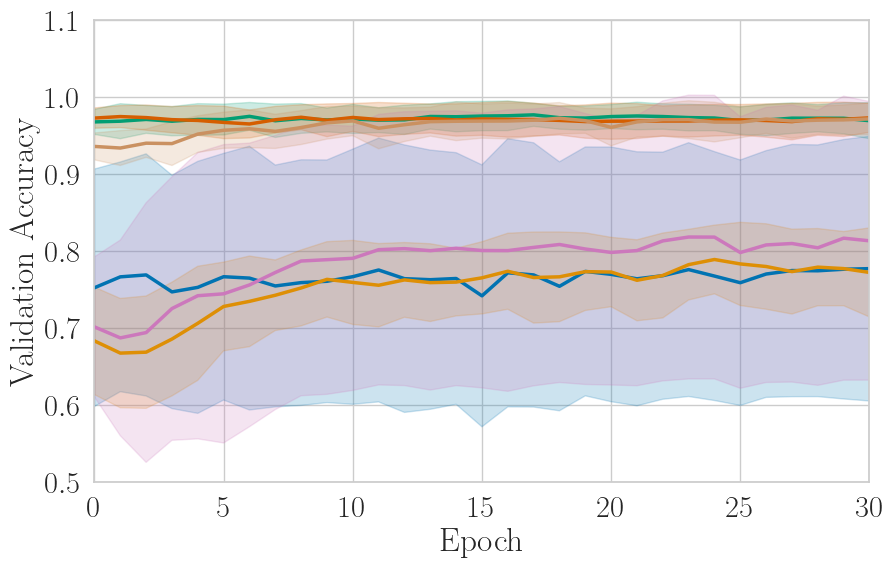}}\\[0pt] 
  \subfloat[MNIST]{
   \label{fig:multipic:MNIST} 
   \includegraphics[width=0.48\linewidth]{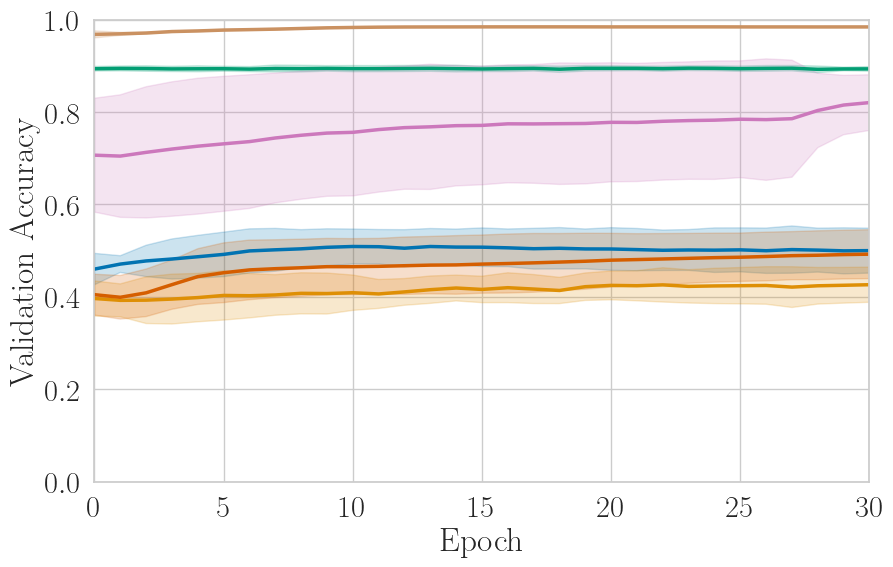}}
   \hfill
  \subfloat[AudioMNIST]{
   \label{fig:multipic:AudioMNIST} 
   \includegraphics[width=0.48\linewidth]{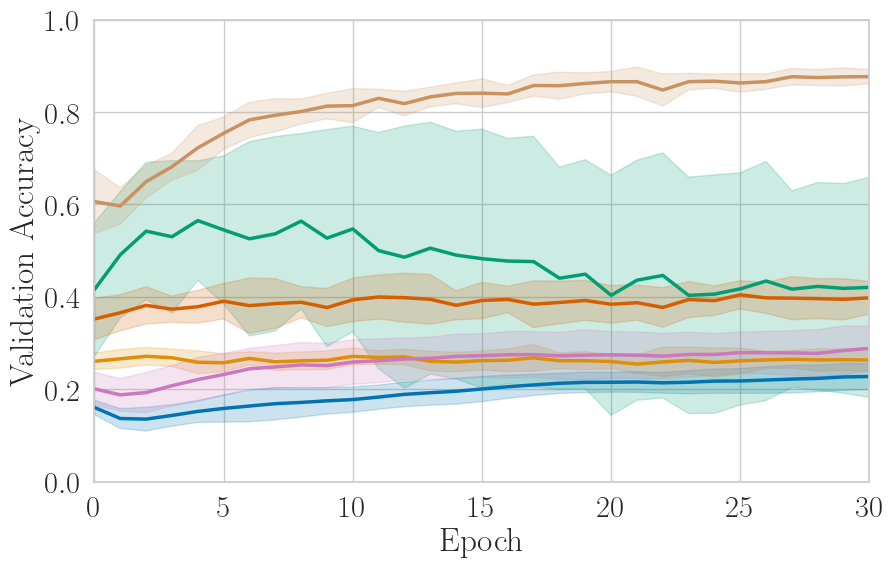}}
 \caption[Validation accuracies for all models and datasets.]{Plots for the validation accuracies for all models over the first 30 epochs. The datasets Banknote Authentication, Breast Cancer Wisconsin, MNIST, and AudioMNIST are displayed.}
 \label{fig:multipic} 
\end{figure*}

In this section, we will first show the performances of the models for each dataset separately, followed by the overall results of the approaches averaged over the datasets. The plots for the validation accuracies can be found in \cref{fig:multipic}. A summary of all test accuracies is presented in \cref{tab:ResultsTable}. 
To check if the AE does not classify as assumed, we conducted tests, the results of which confirmed this assumption.
Additionally, we conducted statistical analyses to check if there are significant differences in the performance of the models. If there are significant differences, we also conducted follow-up tests to analyze the pairwise differences. A significance level of $\alpha=0.05$ was assumed. For the sake of simplicity, only the pairwise differences between our model and the others are stated. The statistical tests and follow-up tests were calculated with SPSS (version 29.0.1.0).

\subsection{Banknote Authentication}\label{subsec:ResultsBanknote}
For the Banknote Authentication dataset, the DQC performed the best with an accuracy of $0.994$, followed by the classical NN on the uncompressed input with $0.991$, and SEQUENT with $0.979$. The VQC with amplitude embedding yielded an accuracy of $0.847$ and therefore performed better than our approach with an accuracy of $0.787$. The classical NN on the compressed input was the worst of the models with $0.699$. 

Subsequently, we checked whether the differences are statistically significant with an ANOVA. Due to the violation of the normality assumption of the repeated measures ANOVA, a Friedman test was calculated, which indicates that the performance differences of the approaches are significant, $\chi^2(5)=45.88, p<0.001$.

Since Friedman's ANOVA revealed significant differences among the approaches, we conducted a follow-up analysis with the Wilcoxon signed-rank test to identify pairwise differences in accuracy. We used adjusted p-values to account for multiple comparisons. Our approach yielded results which are significantly worse than DQC ($p<0.001$), SEQUENT ($p=0.028$), and the NN with uncompressed input ($p=0.001$). It is worth emphasizing that the pairwise comparison of our approach with the VQC with amplitude embedding is not significant ($p>0.999$). The same is true for the NN on the compressed input ($p>0.999$).

\subsection{Breast Cancer Wisconsin}\label{subsec:ResultsBreastCancer}
For the Breast Cancer Wisconsin dataset, the classical NN on the uncompressed input performed best with an accuracy of $0.974$. This performance is closely followed by DQC and SEQUENT with an accuracy of $0.972$ for each. VQC with amplitude embedding yielded a performance of $0.849$, followed by the classical NN on the compressed input with $0.833$. Our approach achieved an accuracy of $0.816$ and therefore performed slightly worse than the VQC with amplitude embedding.

We then checked if the differences are statistically significant. Due to the violation of the normality assumption of the repeated measures ANOVA, a Friedman test was calculated, which indicates that the performance differences of the approaches are significant, $\chi^2(5)=31.72, p<0.001$.

Since Friedman's ANOVA revealed significant differences among the approaches, we conducted a follow-up analysis with the Wilcoxon signed-rank test to identify pairwise differences in accuracy. We used adjusted p-values to account for multiple comparisons. Our model performed significantly worse than DQC ($p=0.042$) and SEQUENT ($p=0.042$). The difference between our approach and the VQC with amplitude embedding is not significant ($p>0.999$). The same is true for the difference between our model and the NN on the compressed input ($p>0.999$) and the NN on the uncompressed input ($p=0.062$).

\subsection{MNIST}\label{subsec:ResultsMNIST}
Also for dataset MNIST, the NN on the uncompressed input obtained the best accuracy with $0.985$. This is followed by DQC with $0.896$ and the classical NN on the compressed input with $0.831$. SEQUENT achieved an accuracy of $0.508$, marginally better than our approach with $0.507$. The worst accuracy of $0.444$ was reached by the VQC with amplitude embedding.

In order to check if the differences are statistically significant, we want to perform a repeated measures ANOVA. Due to limited sphericity ($p<0.001$), Greenhouse-Geiser corrected values are reported ($\epsilon<0.75$). The repeated measures ANOVA indicates that there is a statistically significant performance difference between the approaches, $F(2.77,25.92)=260.67, p<0.001$.

Since the repeated measures ANOVA revealed significant differences among the approaches, we conducted a follow-up analysis to identify pairwise differences in accuracy. We used Bonferroni correction for multiple comparisons. Our model performed significantly worse than DQC ($p<0.001$), the classical NN on the compressed input ($p<0.001$), and the NN on the uncompressed input ($p<0.001$). The difference between our approach and the VQC with amplitude embedding is not significant ($p=0.340$). In addition, the difference between our approach and SEQUENT is not significant ($p>0.999$).

\subsection{AudioMNIST}\label{subsec:ResultsAudioMNIST}
\begin{figure}[hbt]
  \centering
  \includegraphics[width=\columnwidth]{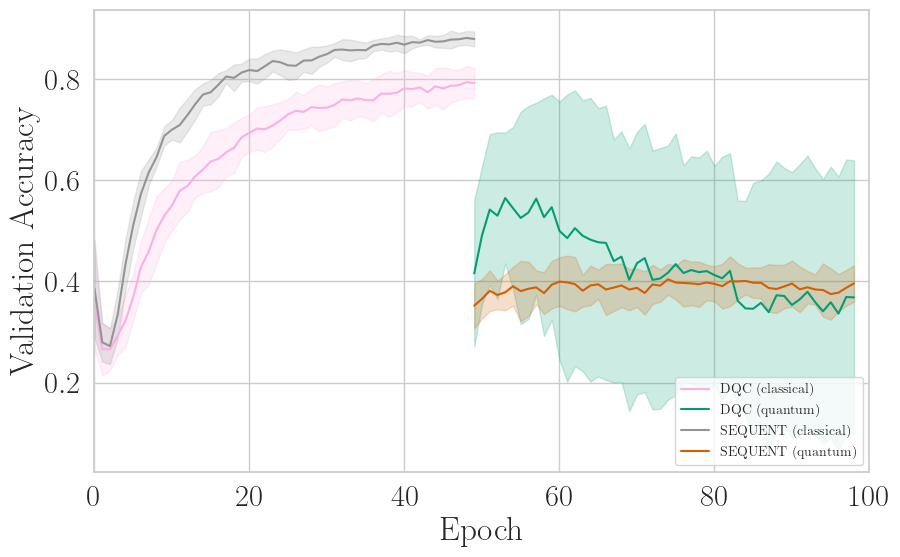}\\
  \caption[Validation accuracies for two-step transfer learning for AudioMNIST.]{Validation accuracies for AudioMNIST. The classical and quantum validation accuracies are shown for the two-step hybrid transfer learning of DQC and SEQUENT.}
  \label{fig:Plot_TL_AudioMNIST}
\end{figure}

For the AudioMNIST dataset, the classical NN on the uncompressed input performed the best with an accuracy of $0.879$, followed by SEQUENT with $0.385$ and DQC with $0.356$. The classical NN on the compressed input achieved an accuracy of $0.298$. The VQC with amplitude embedding yielded a result of $0.257$ and was therefore better than our approach with an accuracy of $0.240$. For DQC and SEQUENT, it can be seen in \cref{fig:Plot_TL_AudioMNIST} that the classical training step achieved validation accuracies of about $0.8$. In the quantum training step, first an expected drop occurred and then only a slight improvement of accuracy can be observed visually for SEQUENT. The validation accuracy of DQC deteriorates.

Subsequently, we checked whether the differences are statistically significant with an ANOVA. Due to the violation of the normality assumption of the repeated measures ANOVA, a Friedman test was calculated. Friedman's ANOVA indicates that the performance differences of the approaches are significant, $\chi^2(5)=33.03, p<0.001$. 

Since Friedman's ANOVA revealed significant differences among the approaches, we conducted a follow-up analysis with the Wilcoxon signed-rank test to identify pairwise differences in accuracy. We used adjusted p-values to account for multiple comparisons. Our approach performed significantly worse than SEQUENT ($p=0.028$) and the classical NN on the uncompressed input ($p<0.001$). No statistically significant difference can be obtained between our model and VQC with amplitude embedding ($p>0.999$), DQC ($p>0.999$), and the classical NN on the compressed input ($p>0.999$).

\begin{table*}[tb]
\centering
\resizebox{\textwidth}{!}{%
\begin{tabular}{l c c c c c c c}
\toprule
Model & AE+VQC (angle) & VQC (amplitude) & DQC & SEQUENT & AE+NN & NN & AE\\ Dataset\\
\midrule
Banknote Authentication & $0.787\pm0.023$ & $0.847\pm0.036$ & $0.994\pm0.004$ & $0.979\pm0.009$ & $0.699\pm0.084$ & $0.991\pm0.008$ & $0.172\pm0.030$ \\
Breast Cancer Wisconsin & $0.816\pm0.114$ & $0.849\pm0.021$ & $0.972\pm0.018$ & $0.972\pm0.016$ & $0.833\pm0.121$ & $0.974\pm0.022$ & $0.016\pm0.024$ \\
MNIST & $0.507\pm0.036$ & $0.444\pm0.033$ & $0.896\pm0.005$ & $0.508\pm0.046$ & $0.831\pm0.042$ & $0.985\pm0.001$ & $<0.001\pm<0.001$ \\
AudioMNIST & $0.240\pm0.028$ & $0.257\pm0.025$ & $0.356\pm0.188$ & $0.385\pm0.036$ & $0.298\pm0.028$ & $0.879\pm0.039$ & $<0.001\pm<0.001$ \\
\midrule
Average & $0.588\pm0.430$ & $0.599\pm0.473$ & $0.805\pm0.480$ & $0.711\pm0.492$ & $0.665\pm0.402$ & $0.957\pm0.084$ & $0.047\pm0.133$ \\
\bottomrule
\end{tabular}
} 
\caption[Test accuracies for all models.]{Test accuracy and $95\%$ confidence interval for all models.\label{tab:ResultsTable}}
\end{table*}

\subsection{Overall Comparison}\label{subsec:ResultsOverall}
Averaged over all datasets, the classical NN on the uncompressed input achieved an accuracy of $0.957$, followed by the DQC with $0.805$ and SEQUENT with $0.711$. The NN on the compressed input gives an accuracy of $0.665$. The VQC with amplitude embedding had a marginally better performance with $0.599$ than our approach with $0.588$. 

In this context, we aim to validate the statistical significance of the differences by conducting a one-way ANOVA (without repeated measures). Due to the violation of the normality assumption of the one-way ANOVA, a Kruskal-Wallis test was performed, which indicates that the performance differences of the approaches are significant, $\chi^2(5)=80.75, p<0.001$. 

Since the Kruskal-Wallis test revealed significant differences among the approaches, we conducted a follow-up analysis with Mann-Whitney tests to identify pairwise differences in accuracy. We used adjusted p-values to account for multiple comparisons. Our approach performed significantly worse than DQC ($p<0.001$) and the NN on the uncompressed input ($p<0.001$). The observed difference between our approach and SEQUENT falls just short of achieving statistical significance ($p=0.052$). Between our approach and the VQC with amplitude embedding no significant difference can be obtained ($p>0.999$). The same is true for the difference between our approach and the NN on the compressed input ($p>0.999$).

\subsection{Discussion}\label{subsec:BA_discussion}
The NN on the uncompressed input, DQC, and SEQUENT achieved better results than our approach, where the former was superior. Across DQC, SEQUENT, and our model the VQC shared the same architecture and numbers of layers. In contrast to DQC and SEQUENT, our compression part (AE) did not classify which can be seen in \cref{tab:ResultsTable}. This leads to the assumption that the classical compression layer plays a pivotal role in the overall performance of these hybrid transfer learning approaches and the role of the VQC itself may be questioned. This assumption is also supported by \cref{fig:Plot_TL_AudioMNIST}, which shows the respective validation accuracies for the two-stage training process of DQC and SEQUENT, exemplary for AudioMNIST. Already after the classical training step, a very high validation accuracy can be seen for both models over both datasets. After this training stage, a drop in the accuracy can be observed because the quantum weights were randomly initialized. The validation accuracy subsequently did not increase any further in the quantum training step, where just the weights of the VQC were optimized - the validation accuracies for DQC and SEQUENT after the complete training are worse than the accuracies after the classical training stage. A possible explanation could be that the extracted features of the pre-processing or compression layer do not contain enough information for further classification, or that the VQC lacks the required power or complexity to maintain the desired results. 

The comparison between the VQC with amplitude embedding and our model did not show any statistically significant difference in performance - making our approach a valid alternative. To test the performance of the AE, a classical NN that uses the reduced input of the AE was introduced. This model achieved good results for Breast Cancer Wisconsin and for MNIST. However, especially for AudioMNIST an accuracy of just $0.298$ was achieved - compared to the NN on the uncompressed input with $0.879$. This suggests that the proposed AE may not have been able to adequately extract the essential information within the latent space. It is hence possible that the architecture is not able to adapt well to specific characteristics of a dataset. Further research is needed to enhance the effectiveness of the AE and come up with specialized architectures for datasets. Additionally, joint training of the AE and the VQC in contrast to sequential training should be considered. Interestingly, the other models (except for the NN on the uncompressed input) also did not exceed 40\% accuracy either. This indicates that the pre-processing of the AudioMNIST dataset may not have been effective. Improving this process could also be a promising direction for future research.

Another limitation of the AE in our approach is the additional effort for the training of the AE, especially compared to the VQC with amplitude embedding. It is also worthwhile to consider other data compression techniques, e.g. the principal component analysis.
As already mentioned, the architecture of our VQC could pose a limitation. It can be beneficial to allow for rotations across all three axes and to increase the number of layers. Other techniques to find the optimal hyperparameters of a model should be considered.
\section{CONCLUSION} \label{sec:conclusion}

In this paper, we introduced an approach to tackle the issues of the current NISQ era. We used the encoder of an AE to reduce the input dimension of a dataset. A versatile architecture was chosen, where each layer halves the number of neurons until the number of neurons in one layer equals the number of output classes. This compressed input is fed to a VQC, which uses angle embedding to map the data from the classical to the Hilbert space. The performance was measured across the four datasets Banknote Authentication, Breast Cancer Wisconsin, MNIST, and AudioMNIST -- ranging from medical to image and audio samples. We then compared our model to other designs: SEQUENT and DQC from the topic of classical to quantum transfer learning, VQC with amplitude embedding as a purely quantum architecture and a purely classical NN on the compressed and uncompressed input. 

Our results suggest that the classification performance in hybrid transfer learning is mainly influenced by the classifying compression layer and that the actual contribution of the VQC may be doubted. This assumption is supported by the fact that the actual training for SEQUENT and DQC took place in the classical training stage. Additionally, these approaches yield better results than models where solely the VQC classifies. 

Even though our model performs worse on average than the hybrid transfer learning models DQC and SEQUENT, it is noticeably better than random guessing and allows for a more transparent and interpretable analysis of the quantum circuit's role in the machine learning task because of the clear distinction between the components. Additionally, our research indicates that our approach with angle embedding on the compressed input is a valid alternative to a VQC with amplitude embedding on the original input. 
\section*{ACKNOWLEDGEMENTS}
This work is part of the Munich Quantum Valley, which is supported by the Bavarian state government with funds from the Hightech Agenda Bayern Plus. This paper was partially funded by the German Federal Ministry for Economic Affairs and Climate Action through the funding program "Quantum Computing -- Applications for the industry" based on the allowance "Development of digital technologies" (contract number: 01MQ22008A).

\bibliographystyle{apalike}
{\small
\bibliography{main}}

\end{document}